\documentclass[11pt]{article}
\usepackage{amssymb}
\usepackage{amsmath,amsthm}  
\usepackage{graphicx}
\usepackage{microtype}
\usepackage[english,francais]{babel}

\newfont{\ensmathquatorze}{msbm10 scaled 1400}
\newfont{\ensmathonze}{msbm10 scaled 1100}
\newfont{\ensmathdix}{msbm10}
\newfont{\ensmathneuf}{msbm10 scaled 833}
\newfont{\ensmathhuit}{msbm10 scaled 694}
\newfam\ensmathfam                        
\textfont\ensmathfam=\ensmathonze        
\scriptfont\ensmathfam=\ensmathdix       
\scriptscriptfont\ensmathfam=\ensmathhuit
\def\ensmf{\fam\ensmathfam\ensmathonze}         

\renewcommand{\leq}{\leqslant}

\def\eqdef{\stackrel{\mbox{\tiny def}}{=}}     
\def\eqlaw{\stackrel{\mbox{\tiny (law)}}{=}}     
\newcommand{\ket}[1]{|\kern.3ex#1\kern.3ex\rangle}
\newcommand{\bra}[1]{\langle\kern.3ex #1 \kern.3ex|}

\newcommand{\mean}[1]{\left\langle #1 \right\rangle} 
\newcommand{\smean}[1]{\langle #1 \rangle} 

\newcommand{\EXP}[1]{{\mbox{\large e}}^{#1}}         

\newcommand{\cotg}{\mathop{\mathrm{cotg}}\nolimits}  
\renewcommand{\min}[2]{\mathop{\mathrm{min}}\nolimits\left( #1 , #2\right)}

\def\RR{{\ensmf R}}                 

\def\D{{\rm d}}                  

\newcommand\antiddots{\mathinner{\mkern2mu\raise1pt\hbox{.}\mkern2mu
\newline \raise4pt\hbox{.}\mkern2mu\raise7pt\hbox{.}\mkern1mu}}

\begin{document}

\selectlanguage{english}

\title{Localization for one-dimensional \\ random  
  potentials with large local fluctuations}

\author{ Tom Bienaim\'e and Christophe Texier }

\date{September 9, 2008}

\maketitle

\hspace{1cm}
\begin{minipage}[t]{13cm}
{\small

Laboratoire de Physique Th\'eorique et Mod\`eles Statistiques,
UMR 8626 du CNRS, 

Universit\'e Paris-Sud, B\^at. 100, F-91405 Orsay Cedex, France.

}
\end{minipage}

\begin{abstract}
  We study the localization of wave functions for one-dimensional
  Schr\"odinger Hamiltonians with random potentials $V(x)$ with short
  range correlations and large local fluctuations such that
  $\int\D{x}\,\smean{V(x)V(0)}=\infty$. A random supersymmetric
  Hamiltonian is also considered.  Depending on how large the
  fluctuations of $V(x)$ are, we find either new energy dependences of
  the localization length, $\ell_\mathrm{loc}\propto{}E/\ln{E}$,
  $\ell_\mathrm{loc}\propto{}E^{\mu/2}$ with $0<\mu<2$ or
  $\ell_\mathrm{loc}\propto\ln^{\mu-1}E$ for $\mu>1$, or
  superlocalization (decay of the wave functions faster than a simple
  exponential).
\end{abstract}

\noindent
PACS numbers~: 72.15.Rn~; 73.20.Fz~; 02.50.-r.





\vspace{0.25cm}

\noindent{\bf Introduction.--}
The phenomenon of Anderson localization \cite{And58} in one dimension
has been widely studied since the pioneering work of Mott \& Twose
arguing that all states are localized in one-dimension
(1d)~\cite{MotTwo61}. This statement was rigorously proven in
Refs.~\cite{GolMolPas77,PasFig78}. A general method to study the
spectral and localization properties of 1d random Hamiltonian was
proposed in Refs.~\cite{AntPasSly81,LifGrePas88}~: let us consider the
one-dimensional Schr\"odinger Hamiltonian
$H=-\frac{\D^2}{\D{}x^2}+V(x)$ where $V(x)$ is a random potential with
short range correlations. We first study the solution of the
stationary Schr\"odinger equation $H\psi(x;E)=E\psi(x;E)$ satisfying
$\psi(0;E)=0$ and $\psi'(0;E)=1$ (differenciation with respect to $x$
is denoted by $'$).  We define the Lyapunov exponent (inverse
localization length $\ell_\mathrm{loc}\equiv1/\gamma$) as the increase
rate~\cite{AntPasSly81,Luc92}
\begin{equation}
  \gamma(E) \eqdef \lim_{x\to\infty}\frac{\D}{\D{x}}
  \bigg\langle \ln\sqrt{\psi(x;E)^2+\frac1{E}\,\psi'(x;E)^2} \bigg\rangle
  \:.
\end{equation}
Averaging $\mean{\cdots}$ is taken over realizations of
the random potential. This definition becomes more clear if the wave
function is parametrized in terms of an envelope and an oscillating
part. We substitute to the couple of functions $(\psi,\psi')$ the
variables $(\theta,\xi)$ according to~:
\begin{align}
  \label{pf1}
  \psi(x;E)  &= \phantom{k\,}\EXP{\xi(x)}\, \sin\theta(x) \\
  \label{pf2}
  \psi'(x;E) &= k\,          \EXP{\xi(x)}\, \cos\theta(x) 
  \:,
\end{align}
where $E=k^2$. We can rewrite the definition of the Lyapunov exponent as
$\gamma(E)=\frac{\D}{\D{x}}\smean{\xi(x)}$.  Therefore the Lyapunov
exponent gives the rate of the exponential increase of the envelope of
the wave function.

At high energy (compared to disorder), oscillations of the wave
function occur on the typical scale $k^{-1}$ and the Lyapunov exponent
is given by~\cite{AntPasSly81,LifGrePas88}\footnote{
  The solution $\psi(x;E)$ of the Cauchy problem exists for any value
  of the energy~; it is used to construct the normalized wave
  functions $\varphi(x)$ of the stationary Schr\"odinger equation on a
  finite interval $[0,L]$ satisfying boundary conditions
  $\varphi(0)=\varphi(L)=0$, what can only be achieved for a discrete
  set of energies (Sturm-Liouville problem). From this scheme we
  expect that the normalized wave functions present the structure
  $\varphi(x)\sim\sin(kx+\theta_0)\,\EXP{-|x-x_0|/\ell_\mathrm{loc}}$.
  Note however that this simple picture neglects the important fact
  that, in the exponential, $\xi(x)$ has large absolute fluctuations
  despite it presents negligible fluctuations relatively to its
  average when $\int\D{x}\,\smean{V(x)V(0)}<\infty$. These
  fluctuations play a very important role since they induce large
  fluctuations of the normalization of the wave function (see \S13.3
  of Ref.~\cite{LifGrePas88})~\cite{TexCom99}.  
%
}
\begin{equation}
  \label{Pastur}
  \gamma(E\to\infty) \simeq \frac{1}{8k^2}\int\D{x}\,\smean{V(x)V(0)}\,\cos2kx
  \:.
\end{equation} 

The question of the present article is to discuss the situation where
the potential presents large fluctuations such that
$\int\D{x}\,\smean{V(x)V(0)}=\infty$, which makes (\ref{Pastur}) 
inapplicable.
The integral of the correlation function may diverge for different
reasons.  First, it may diverge due to long range correlations. This
corresponds to nonstationary potentials. The localization for self
affine potentials such that $\smean{[V(x)-V(0)]^2}\propto|x|^{2h}$,
with the Hurst exponent $h>0$, was studied in
Refs.~\cite{DeMLyr98,RusKanBunHav01,Luc05} (the case $h=1/2$
corresponds to the Brownian case, partly studied in
Ref.~\cite{JosJay98}). This will not be the question of interest in
the present article.  Another reason for the divergence
$\int\D{x}\,\smean{V(x)V(0)}=\infty$ is for a potential with {\it
  short range} correlations and {\it large local} fluctuations.  This
is the case on which we will focus here.

For simplicity we consider a random potential $V(x)$ uncorrelated at
different positions (vanishing correlation length).  A model that
realizes these conditions is the following random potential
\begin{equation}
  \label{potential}
  V(x) = \sum_n v_n  \, \delta(x-x_n)
  \:,
\end{equation}
where the weights $v_n$ are chosen to be independent and identical random
variables distributed according to a distribution with power law tail
\begin{equation}
  \label{powerlaw}
  p_1(v) \propto \frac1w\left|\frac{w}{v}\right|^{1+\mu}
  \hspace{0.5cm}\mbox{for } v \to \pm\infty
\end{equation}
with $\mu>0$. Here $w$ is a scale for the weights to make the
argument of the tail dimensionless.  The positions
of impurities $x_n$ are also chosen to be independent random variables
uniformly distributed with a finite density $\rho$.
When $\mu\leq2$ the second moment diverges $\smean{v_n^2}=\infty$, as
well as the correlation function of the potential (\ref{potential}) since
$\smean{V(x)V(x')}=\rho\,\smean{v_n^2}\,\delta(x-x')$.

Some exact results have been obtained for a tight binding Hamiltonian
with random on-site potential (Anderson model) distributed according
to a Cauchy law \cite{Llo69,Tho72,Ish73} (\S10.3 of
Ref.~\cite{LifGrePas88} or Ref.~\cite{Luc92}).  
Note also that
fluctuations of the envelope of the wave function and conductance
statistics for this discrete model were studied for power law disorder
in the recent works \cite{DeyLisAlt01,TitSch03}.
The Anderson model (AM) can be mapped \cite{Luc92} onto the problem we are
interested in here for {\it fixed} impurity positions and with
$\mu=1$. Potentials for fixed and random impurity positions share some
features, however, when impurities of random weights form a lattice,
$x_n=n/\rho$, the trace of the lattice remains for arbitrary large
energies (band edges remain at $k_n=n\pi\rho$). This makes the
definition of a high energy regime less convenient.

\vspace{0.25cm}

\noindent{\bf Ricatti variable.--}
We follow the ideas introduced in Ref.~\cite{FriLlo60} in order to
study the spectrum, and apply them to the localization problem.  Let
us introduce the Ricatti variable
$z(x)\eqdef\frac{\psi'(x)}{\psi(x)}$.  From the Schr\"odinger equation
we see that it obeys a Langevin equation $z'=-E-z^2+V(x)$ for initial
condition $z(0)=\infty$. The distribution $T(z;x)$ of the Ricatti
variable obeys the integro-differential equation
\begin{equation}
  \frac{\partial}{\partial x}T(z;x)
  = \frac{\partial}{\partial z}\left[(E+z^2)T(z;x)\right]
  + \rho\int\D v\,p(v)\,\left[T(z-v;x)-T(z;x)\right]
  \:.
\end{equation}
The first term in the right hand side is the drift term coming from
the force $-(E+z^2)$ and the second a jump term originating from the
random potential (\ref{potential}). For $x$ sufficiently large, the
distribution reaches a limiting distribution $T(z)$ for a steady
current~\cite{LifGrePas88,Luc92}. Current of the Ricatti variable
through $\RR$ gives the number of zeros of the wave function per unit
length. This is also the integrated density of states (IDoS) per unit
length $N(E)$, therefore
\begin{equation}
  \label{eqInteg}
  N(E) = (E+z^2)T(z)  - \rho \int\D v\,p(v)
  \int_{z-v}^z\D z'\,T(z')
  \:.
\end{equation}
Imposing normalization of the solution of this integral equation gives
the IDoS. Knowing the limiting distribution $T(z)$, the Lyapunov
exponent can be obtained from~\cite{LifGrePas88} $\gamma=\smean{z}$.
Since $T(z\to\pm\infty)\simeq{}N(E)/z^2$, in order to deal with well
defined integral it is understood that calculation of the Lyapunov
exponent involves the antisymmetric part of the
distribution~:~$\gamma=\int\D{z}\,z\,\frac12[T(z)-T(-z)]$).

Let us study the high energy Lyapunov exponent. For that purpose we
solve the integral equation (\ref{eqInteg}) by perturbation starting
from the solution for $V(x)=0$. In the absence of disorder
($p(v)=\delta(v)$) we have $T_0(z)=\frac1\pi\frac{k}{z^2+k^2}$. We
expand the distribution $T(z)=T_0(z)+T_1(z)+\cdots$ in powers of the
density $\rho$, as well as the IDoS.  Then eq.~(\ref{eqInteg}) is
solved recursively order by order.  We easily obtain $T_1(z)$ from
which we deduce
\begin{equation}
  \gamma(E=k^2\to\infty) \simeq \frac\rho\pi \int\D z\, \frac{z}{z^2+k^2}
  \int\D v\,p(v)\,
  \left[ \arctan\frac{z}{k} - \arctan\frac{z-v}{k} \right] 
  \:.
\end{equation}
This gives the general formula
\begin{equation}
  \label{RES1}
  \gamma(k^2) \simeq \frac\rho2
  \mean{\ln\left[1+\left(\frac{v}{2k}\right)^2\right]}_v
  \:,
\end{equation}
where the averaging is now taken over the $\delta$-peak weights~$v_n$.  
This is the first term of a ``concentration expansion'' that can be
systematically developed~\cite{LifGrePas88} (eq.~(\ref{RES1}) was
derived in the \S10.4 of this latter reference for non random weights
$v_n$. Additional averaging in eq.~(\ref{RES1}) follows from the property
of additivity of the variable $\xi(x)$).

\vspace{0.25cm}

\noindent{\bf New energy dependences.--}
We first consider the high energy Lyapunov exponent, $\sqrt{E}=k\gg\rho,\,w$,
when the weights are distributed according to (\ref{powerlaw}).  We
write $p_1(v)=\frac1wf(v/w)$ where $f(y)$ is a dimensionless symmetric
function such that $f(y\to\pm\infty)\simeq{}C\,|y|^{-1-\mu}$.  We
divide 
the integral
$\gamma\simeq\rho\int_0^\infty\D{y}\,f(y)\,\ln[1+(\frac{w}{2k}y)^2]$
into three parts~: $\gamma\sim\rho\big[
(\frac{w}{2k})^2\int_0^1\D{y}\,f(y)\,y^2
+(\frac{w}{2k})^2C\int_1^{2k/w}\D{y}\,y^{1-\mu}
+2C\int_{2k/w}^\infty\D{y}\,y^{-1-\mu}\ln(\frac{w}{2k}y)\big]$.

\noindent$\bullet$ For $\mu>2$ the Lyapunov exponent is dominated by
smallest $y$ ($\lesssim1$). We obtain $\gamma\propto\rho(\frac{w}{k})^2$ that
corresponds to expand the logarithm of
eq.~(\ref{RES1}) for small $v$. This is the result of
eq.~(\ref{Pastur})~:
$\gamma(k^2)\simeq\frac1{8k^2}\rho\,\smean{v^2}$.

\noindent$\bullet$ For $\mu=2$, eq.~(\ref{Pastur}) cannot be used
since $\smean{v^2}=\infty$. The integral giving the Lyapunov exponent is
dominated by the intermediate scale $1\lesssim{}y\lesssim{}k/w$~:
\begin{eqnarray}
  \label{lyap2}
    \gamma(k^2) \propto  \rho\left(\frac{w}{k}\right)^2\ln\left(\frac{2k}{w}\right)
  \:.
\end{eqnarray}

\noindent$\bullet$ For $0<\mu<2$, the Lyapunov exponent is
dominated by largest $y$ ($\gtrsim{}k/w$). We obtain
\begin{eqnarray}
  \label{lyap1}
  \gamma(k^2) 
    \propto  \rho\left(\frac{w}{k}\right)^\mu  
  \:.
\end{eqnarray}
%
The numerical dimensionless prefactors depend on the precise form of the
distribution and not only on its tail.

The fluctuations of the random weights can be further increased by considering
a distribution with tail~:
\begin{equation}
  \label{loglaw}
  p_2(v) \propto \frac1{|v|\ln^{1+\mu}\left|\frac{v}{w}\right|}
  \hspace{0.5cm}\mbox{for }  v \to \pm\infty
\end{equation}
for $\mu>0$.  When $\mu>1$ we find that the Lyapunov exponent decays logarithmically
with energy
\begin{equation}
  \gamma(k^2) \sim \frac{\rho}{\ln^{\mu-1}(\frac{k}{w})}
  \:.
\end{equation}
The case $0<\mu\leq1$ is discussed in the next section.

\vspace{0.25cm}

\noindent{\bf Superlocalization.--}
On the other hand, for distribution (\ref{loglaw}) with $\mu\leq1$,
not only the second moment diverges $\smean{v^2}=\infty$, but the
expression (\ref{RES1}) shows that the Lyapunov exponent diverges as
well~: $\gamma=\infty$. This indicates that the logarithm of the
envelope of the wave function, $\xi(x)$, presents different scaling
properties with $x$. In order to analyze this, we remark that the
variable $\xi(x)$ is constant between two impurities and makes a jump
$\Delta\xi_n\eqdef\xi(x_n^+)-\xi(x_n^-)\sim\ln|v_n|$ across the
impurity $n$ (see below, the section on numerics). Therefore $\xi(x)$
behaves as the sum of $N\sim\rho{}x$ independent variables, each
distributed according to a power law distribution
$p(\Delta\xi)\propto\Delta\xi^{-1-\mu}$.  Using well known results
(recalled in appendix A) we obtain
\begin{eqnarray}
  \label{unconv1}
  \xi(x) &\sim& (\rho x)^{1/\mu}
    \hspace{1.5cm}\mbox{for }0<\mu<1 \\
  \label{unconv2}
  &\sim& (\rho x)\ln(\rho x)
    \hspace{1cm}\mbox{for }\mu=1
  \:.
\end{eqnarray}
The envelope of the wave function presents a decay faster than a
simple exponential.  This phenomenon is called {\it superlocalization}
and has been recently studied for a discrete model in
Ref.~\cite{BooLuc07}\footnote{ Note also that such superlocalization
  $\xi(x)\sim{x}^{1+h/2}$ occurs for self affine random potentials
  characterized by long-range correlations
  $\smean{[V(x)-V(0)]^2}\propto|x|^{2h}$ with $h>0$ \cite{Luc05}.  }.
Characterization of the localization properties cannot be limited to
the typical behaviours (\ref{unconv1},\ref{unconv2}) since the
variable $\xi(x)$ presents large fluctuations. Its distribution is
characterized by a power law tail
\begin{equation}
  \label{PdeXi}
  \mathcal{P}(\xi;x)\propto1/\xi^{1+\mu}
\end{equation}
with the same exponent as the one involved in the distribution of the weights.

\vspace{0.15cm}

\noindent{\it Conductance.--}
We give another interpretation of the previous result in terms of the
conductance of a finite disordered interval of length $L$. The
dimensionless conductance is equal to the transmission probability
(Landauer formula), and presents the same exponential decay as the
square of the wave function modulus.  Therefore we can write
$g\sim\EXP{-2\xi(L)}$ (a more precise definition of the reflection
coefficient within the phase formalism can be found in
Ref.~\cite{AntPasSly81}).

Let us first recall some well known results valid for a potential with
finite local fluctuations. At high energy, when (\ref{Pastur}) holds,
$\xi(x)$ behaves like a Brownian motion with drift\footnote{The fact
  that drift and the Gaussian fluctuations involve the same parameter
  is refered to as ``{\it single parameter scaling}''
  \cite{AbrAndLicRam79} (see also \cite{DeyLisAlt01})~; it holds only
  at high energy since it relies on the decoupling between a fast
  variable (the phase $\theta(x)$ introduced above) and the slow
  variable $\xi(x)$.}  \cite{AntPasSly81}~:
$\xi(x)\eqlaw\gamma\,x+\sqrt{\gamma}\,W(x)$, where $W(x)$ is a Wiener
process\footnote{a normalized free Brownian motion such that
  $\mean{W(x)}=0$ and $\mean{W(x)W(x')}=\min{x}{x'}$.}.  It follows
that the distribution of the logarithm of the conductance is Gaussian
$\Pi(\ln{g})\simeq\frac1{\sqrt{8\pi\gamma\,L}}\exp-\frac1{8\gamma\,L}(\ln{g}+2\gamma{}L)^2$
\cite{Pen94}. The typical value of the conductance is
$g_\mathrm{typ}\sim\EXP{-6\gamma{}L}$ (while
$(\ln{}g)_\mathrm{typ}\sim-2\gamma{}L$) however fluctuations of the
logarithm are associated to a much larger scale
$g_\mathrm{fluct}\sim\EXP{-2\sqrt{\gamma{}L}}$.

Distribution of the conductance in the Anderson model with power law
disorder has been studied in Ref.~\cite{TitSch03} where some power law
distribution of the conductance was obtained for~$g\to0$.


In the superlocalization regime, the behaviour (\ref{unconv1}) is
associated with a decay of the conductance $g\sim\EXP{-L^{1/\mu}}$.
The distribution (\ref{PdeXi}) can be related to the conductance
distribution~:
\begin{equation}
  \Pi( \ln g) \underset{g\to0}{\sim} \frac1{|\ln g|^{\mu+1}}
\end{equation}
for $0<\mu\leq1$.

\vspace{0.25cm}

\noindent{\bf Localization for supersymmetric Hamiltonian.--}
We consider another class of random Hamiltonians with the so-called
supersymmetric structure
\begin{equation}
  \label{Hsusy}
  H = -\frac{\D^2}{\D x^2} + \phi(x)^2 + \phi'(x)
  \hspace{0.5cm}\mbox{with }   \phi(x) = \sum_n \eta_n  \, \delta(x-x_n)
  \:,
\end{equation}
where $\eta_n$ are dimensionless uncorrelated weights, each
distributed according to a distribution $p(\eta)$.  This Hamiltonian
is interesting since it presents rather different spectral and
localization properties (in particular it leads to a delocalization
transition as $E\to0$). It is related to several other problems as
well. For example it is the square of a Dirac Hamiltonian with a
random mass $\phi(x)$, introduced in various contexts of condensed
matter physics~; the problem can also be related to
classical diffusion in a random force field (see
Refs.~\cite{LifGrePas88,BouComGeoLeD90,ComTex98} for a review).  
We can follow the same strategy~: the
Ricatti variable $z=\frac{\psi'}{\psi}-\phi$ obeys the Langevin type
equation $z'=-E-z^2-2z\phi(x)$ with multiplicative noise. Limiting
distribution of the Ricatti variable for a steady current $-N(E)$ obeys the
integral equation
\begin{equation}
  \label{eqIntegSusy}
  N(E) = (E+z^2)T(z)  + \rho \int\D\eta\,p(\eta)
  \int_{z}^{z\mathrm{e}^{2\eta}}\D z'\,T(z')
  \:.
\end{equation}
The Lyapunov exponent is now given by $\gamma=\mean{z}+\mean{\phi}$.
Following the same perturbative approach as before, we obtain
\begin{equation}
   \label{RESsusy}
  \gamma(E\to\infty) \simeq \rho
  \mean{\ln\cosh\eta}_\eta
  \:.
\end{equation}
For the supersymmetric Hamiltonian, when the average exists, the
Lyapunov exponent reaches a finite value at high energy (in contrast
with the decrease of the Lyapunov exponent for the Schr\"odinger
Hamiltonian)~: for $|\eta_n|\ll1$ it takes the form
$\gamma(E\to\infty)\simeq\frac12\int\D{x}\,\smean{\phi(x)\phi(0)}$.

Let us consider a power law distribution of weights
$p(\eta)\propto1/|\eta|^{1+\mu}$. We see from (\ref{RESsusy}) that we
have $\gamma=\infty$ in this case for $\mu\leq1$. The reason is
similar to the one discussed in the previous paragraph for the
Schr\"odinger equation with weights distributed according to
(\ref{loglaw}). Here the variable $\xi(x)$ jumps by
$\Delta\xi_n\sim|\eta_n|$ across the impurity (see below). Therefore,
for the supersymmetric case, a power law distribution of the weights
leads to the superlocalization,
eqs.~(\ref{unconv1},\ref{unconv2},\ref{PdeXi}).

\vspace{0.25cm}

\noindent{\bf Numerical calculations.--} 
We can easily study the evolution of the phase and envelope variables
(\ref{pf1},\ref{pf2}) numerically.  We denote by
$\theta_n^\pm\eqdef\theta(x_n^\pm)$ and $\xi_n^\pm\eqdef\xi(x_n^\pm)$
the value of the phase and the envelope just before and right after
the $n$-th $\delta$-peak.  Between two impurities we have
$\theta_{n+1}^--\theta_n^+=k\ell_n$ and $\xi_{n+1}^--\xi_n^+=0$. The
length $\ell_n=x_{n+1}-x_n$ denotes the distance between consecutive
impurities. It is distributed according to a Poisson
law~$p(\ell)=\rho\,\EXP{-\rho\ell}$.  The evolution of the random
variables across an impurity depends on the form of the random
potential. We introduce the
notation~$\Delta\xi_n=\xi_n^+-\xi_n^-=\xi_{n+1}^--\xi_n^-$.

\noindent$\bullet$
For the {\it Schr\"odinger Hamiltonian} with potential (\ref{potential})
phase evolution is given by 
$\cotg\theta_n^+-\cotg\theta_n^-=\frac{v_n}{k}$ and evolution of
the envelope by
$
\Delta\xi_n=\ln\frac{\sin\theta_n^-}{\sin\theta_n^+}
=\frac12\ln[1+\frac{v_n}{k}\sin2\theta_n^-+\frac{v_n^2}{k^2}\sin^2\theta_n^-]
$.

\noindent$\bullet$
For the {\it supersymmetric Hamiltonian} (\ref{Hsusy}) we have 
$\tan\theta_n^+=\EXP{2\eta_n}\tan\theta_n^-$ and 
$
\Delta\xi_n=\frac12\ln\frac{\sin2\theta_n^-}{\sin2\theta_n^+}
=\frac12\ln[\EXP{2\eta_n}\sin^2\theta_n^-+\EXP{-2\eta_n}\cos^2\theta_n^-] 
$.

\noindent
IDoS is given by $N(E)=\lim_{L\to\infty}\frac{\theta(L)}{L\pi}$ and the
Lyapunov exponent by $\gamma(E)=\lim_{L\to\infty}\frac{\xi(L)}{L}$.

\begin{figure}[!ht]
  \centering
  \includegraphics[scale=0.45]{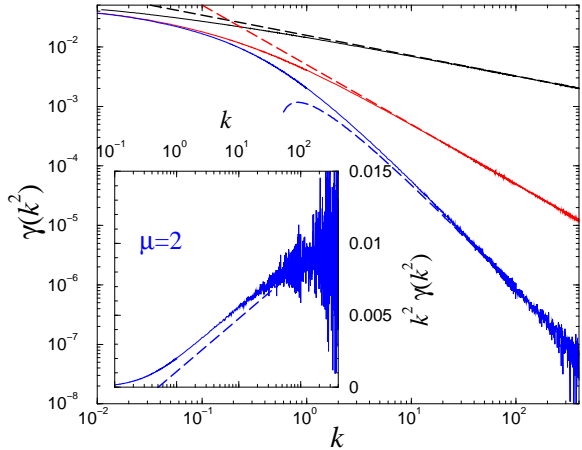}
  \caption{\it For the Schr\"odinger Hamiltonian~: Lyapunov exponent
    as a function of the energy for weights $v_n$ distributed
    according to (\ref{ppl}). Numerical results (continuous lines) are
    compared with high energy expressions (dashed lines) derived in
    the text (no fit) for $\mu=0.34$, $1$ and~$2$.  Other parameters
    are $w=1$, $\rho=0.01$ and number of impurities $N=10^6$.  Inset~:
    $E\gamma(E)$ is plotted in semilog scale for $\mu=2$ in order to
    check  its logarithmic behaviour (dashed line corresponds to
    eq.~(\ref{lyap2})). }
  \label{fig:GamE}
\end{figure}

To be precise we consider a specific distribution with power law tail~:
\begin{equation}
  \label{ppl}
  p_1(v) =
  \frac{\mu\,\left|\frac{v}{w}\right|^{\mu-1}}
       {\pi w\,\big(1+\left|\frac{v}{w}\right|^{2\mu}\big)}
  \:.
\end{equation}
This choice has the advantage that it is very easy to simulate since
the cumulative distribution is straightforwardly obtained.  Using
eq.~(\ref{RES1}), we get the high energy Lyapunov exponent~:
$\gamma\simeq\frac1{\sin(\frac{\pi\mu}2)}\rho(\frac{w}{2k})^\mu$ for
$0<\mu<2$ and
$\gamma\simeq\frac2\pi\rho(\frac{w}{2k})^2\ln(\frac{2k}{w})$ for
$\mu=2$ (both expressions are valid for $k\gg\rho,\,w$).  The case
$\mu=1$ corresponds to a Cauchy law. The right hand side of
eq.~(\ref{RES1}) can be computed easily in this case and we obtain the
expression $\gamma\simeq\rho\ln(1+\frac{w}{2k})$ valid in a broader
range of energy (for $k\gg\rho$ but $w$ arbitrary)~; at high energy we
recover the known energy dependence $\gamma\simeq\rho\frac{w}{2k}$ (it
can be recovered from discrete models~\cite{Tho72,Luc92}).  These
expressions are compared to the numerical results on
figure~\ref{fig:GamE} and work perfectly well.

\begin{figure}[!ht]
  \centering
  \includegraphics[scale=0.45]{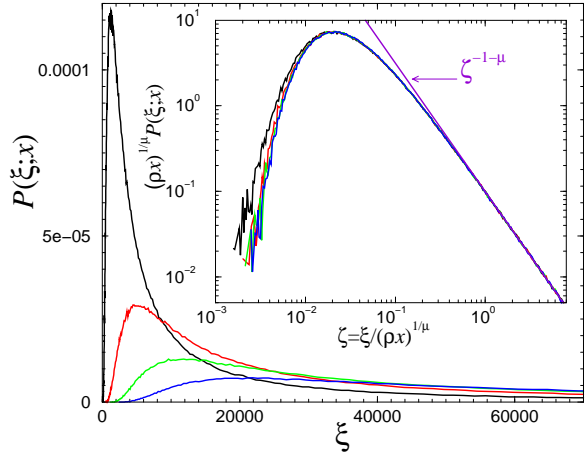}
  \caption{\it Superlocalization for the supersymmetric Hamiltonian.
    Distribution of the variable $\xi(x)$ for
    different values of $x=L/4$, $L/2$, $3L/4$ and $L$.
    Parameters are~: $k=10$, $\rho=1$, $\Lambda=0.1$ (typical scale
    for weights $\eta_n$) and $\mu=0.5$. 
    Number of impurities is $N=10^6$ and $L=1000$. 
    In the inset, straight line corresponds to eq.~(\ref{PdeXi}).}
  \label{fig:PdeXi}
\end{figure}

Next we analyze the superlocalization regime~: we consider
the supersymmetric Hamiltonian for weights $\eta_n$ distributed
according to a law similar to (\ref{ppl}) for $0<\mu<1$. The
distribution $\mathcal{P}(\xi;x)$ is plotted for different values of
$x$ on figure~\ref{fig:PdeXi}. In the inset the axes are rescaled in
order to check that, according to (\ref{unconv1}), the distribution
has the form
\begin{equation}
  \mathcal{P}(\xi;x) \simeq \frac1{(\rho{}x)^{1/\mu}}\:
  \varpi\left(\frac\xi{(\rho{}x)^{1/\mu}}\right)
  \:,
\end{equation}
where $\varpi(\zeta)$ is a dimensionless function. After rescaling we
see that the four curves corresponding to different values of $x$
perfectly collapse onto each other, apart for small deviations
corresponding to the smallest values of $\xi$ and $x$.  Finally we
check that the tail of the distribution is indeed a power law,
eq.~(\ref{PdeXi})~: in the inset of figure~\ref{fig:PdeXi} rescaled
distributions are plotted on a log-log scale with
$\varpi(\zeta)\propto\zeta^{-1-\mu}$. The agreement seems excellent.

\vspace{0.25cm}

\noindent{\bf Conclusion.--} 
We have analyzed the high energy localization length for random
potentials with short range correlations and large local fluctuations
such that the well-known result (\ref{Pastur}), leading to
$\ell_\mathrm{loc}\propto{}E$, is not valid.  
We have studied localization for potentials made of superposition of
$\delta$-peaks. Performing a concentration expansion, we have obtained
two general high energy formulae for the Lyapunov exponent~:
$\gamma(E)\simeq\rho\smean{\ln[1+\frac{v^2}{4E}]}_v$
\cite{LifGrePas88} for the Schr\"odinger Hamiltonian and
$\gamma(E)\simeq\rho\smean{\ln\cosh\eta}_\eta$ for the supersymmetric
Hamiltonian.  These formulae have been used to analyze the case of
potential with large local fluctuations.  

For the Schr\"odinger case, we have shown the relation between the
distribution of the weights of the $\delta$-peaks and the energy decay
of the Lyapunov exponent (inverse localization length
$\ell_\mathrm{loc}$).  Sufficiently large fluctuations of the weights,
such that $\smean{v_n^2}=\infty$, lead to a stronger localization
effect characterized by an increase of $\ell_\mathrm{loc}$ with energy
slower than linear.  These results are summarized in 
table~\ref{tab:summary}.

\begin{table}[!ht]
  \centering
  \begin{tabular}{lll}
     Potential distribution & & Localization length \\
     \hline\hline
     $\smean{v_n^2}<\infty$ & & $\ell_\mathrm{loc}\propto{}E$
     \ \hspace{1cm}\cite{LifGrePas88}\\
     \hline
     $p(v\to\pm\infty)\propto1/|v|^{\mu+1}$ 
         & $\mu=2$  & $\ell_\mathrm{loc}\propto{}E/\ln{}E$   \\
         & $0<\mu<2$ & $\ell_\mathrm{loc}\propto{}E^{\mu/2}$ \\
     \hline
     $p(v\to\pm\infty)\propto\frac1{|v|}\ln^{-1-\mu}\left|\frac{v}{w}\right|$
        & $\mu>1$ & $\ell_\mathrm{loc}\propto\ln^{\mu-1}E$ \\
     \ \hspace{2cm}superlocalization 
        &
          $\left\{\begin{array}{l} \mu=1 \\ 0<\mu<1 \end{array}\right.$
        & $\begin{array}{l} \xi(x)\sim x\ln x \\
          \xi(x)\sim{}x^{1/\mu} \end{array}$
        \\
    \hline\hline
  \end{tabular}
  \caption{\it Energy dependence of the localization length for the
    Schr\"odinger Hamiltonian with random potential
    $V(x)=\sum_nv_n\,\delta(x-x_n)$ for different broad distributions of the
    weights $v_n$.} 
  \label{tab:summary}
\end{table}

%

The understanding of fluctuations of the variable $\xi(x)$ ({\it i.e.}
of the localization length) plays a major role to analyze universal
statistical properties of Wigner time delay \cite{TexCom99}.  It would
be an interesting issue to study how the statistics of Wigner time
delay are affected by the unconventional localization properties
analyzed here.

\vspace{0.25cm}

\noindent{\bf Acknowlegments.--} 
We thank Alain Comtet and Satya Majumdar for interesting remarks and
Jean-Marc Luck for bringing to our attention
Refs.~\cite{Luc05,BooLuc07}.

\vspace{0.25cm}

\noindent{\bf Appendix A.--}
We recall well known results on the distribution of the sum of
independent and identically distributed ({\it i.i.d}) random
variables. Let us consider $N$ {\it i.i.d.} positive  variables $y_n$ and their
sum $Y_N=\sum_{n=1}^Ny_n$. If $\smean{y_n^2}<\infty$ the statistical
properties of $Y_N$ are given by central limit theorem for
$N\to\infty$~: Gaussian distribution centered on
$\smean{Y_N}=N\smean{y}$ of variance
$\smean{Y_N^2}_c=\smean{Y_N^2}-\smean{Y_N}^2=N\smean{y^2}_c$. If the
distribution of the $y_n$'s presents a power law tail
$p(y)\propto1/y^{\mu+1}$ with $0<\mu\leq2$ such that
$\smean{y_n^2}=\infty$, the situation is different~:

\noindent$\bullet$ For $0<\mu<1$ all moments of $y_n$ diverge. Let us
consider the characteristic function $g(p)=\smean{\EXP{-py}}$. We can
write $g(p)=1-\int_0^\infty\D{}y\,(1-\EXP{-py})p(y)$ from which we see
that $g(p\to0)\simeq1-C\,p^\mu$ where $C$ is some constant related to
the prefactor of the power law tail of $p(y)$. The characteristic
function for $Y_N$ reads $G_N(p\to0)\simeq\EXP{-NC\,p^\mu}$. This
shows that the related distribution $P_N(Y)$ presents a similar power
law tail and involves the typical scale $Y_N\sim{}N^{1/\mu}$.

\noindent$\bullet$ For $\mu=1$. A similar analysis gives
$g(p\to0)\simeq1-C\,p\ln1/p$ and therefore $Y_N\sim{}N\ln{N}$.

\noindent$\bullet$ For $1<\mu<2$ the first moment is finite
$\smean{Y_N}=N\mean{y}$ however fluctuations are larger than in the normal case
$\smean{Y_N^2}_c\sim{}N^{2/\mu}$.

\noindent$\bullet$ For $\mu=2$ fluctuations are 
$\smean{Y_N^2}_c\sim{}N\ln{N}$.

\noindent$\bullet$ For $\mu>2$, central limit theorem applies.


\end{document}